\documentclass[a4paper]{article}
\usepackage[left=20mm,right=20mm,top=20mm,bottom=25mm]{geometry}
\usepackage[affil-it]{authblk}
\usepackage{amssymb}
\usepackage{amsmath}
\usepackage{mathtools}
\usepackage[caption=false]{subfig}
\usepackage{dcolumn}
\usepackage{braket}
\usepackage{hyperref}
\usepackage{microtype}
\usepackage{tikz}

\newcommand{\abs}[1]{\lvert #1 \rvert}

\DeclareMathOperator{\tr}{tr}

\hypersetup{
 colorlinks,
 linkcolor={blue!50!black},
 citecolor={blue!50!black},
 urlcolor={blue!50!black}
}

\begin{document}

\title{Looking for chiral anomaly in $K \gamma \to K \pi$ reactions}

\author[1,2,3]{M. I. Vysotsky}
\author[1,3]{E. V. Zhemchugov}
\affil[1]{A. I. Alikhanov Institute for Theoretical and Experimental Physics,
117218 Moscow, Russia}
\affil[2]{Moscow Institute for Physics and Technology, 141700 Dolgoprudny,
Moscow Region, Russia}
\affil[3]{Moscow Engineering Physics Institute, 115409 Moscow, Russia}

\date{}

\maketitle

\begin{abstract}
In an experiment currently being performed at the Institute for High Energy
Physics, Serpukhov, Russia, a beam of charged kaons is directed on a copper
target. In the electromagnetic field of the target nuclei, two reactions occur:
$K^+ \gamma \to K^+ \pi^0$ and $K^+ \gamma \to K^0 \pi^+$. A peculiar
distinction between these two reactions is that there is a chiral anomaly
contribution in the former reaction, but not in the latter.  This contribution
can be directly seen through comparison of the cross sections of these reactions
near the threshold. We derive expressions for these cross sections taking into
account the anomaly and the contribution of the lightest vector mesons.
\end{abstract}

\section{Introduction}

Although the subject of the present paper is the analysis of $K^+ \gamma \to K
\pi$ amplitudes near the threshold, let us begin by reminding an analogous
consideration performed in the literature for the $\pi^+ \gamma \to \pi^+ \pi^0$
amplitude.

$SU(2)_L \times SU(2)_R$ chiral symmetry holds due to the smallness of light
quark masses: $m_{u, d} \ll \Lambda_\text{QCD}$. The anomaly in the divergence
of the axial vector current allows us to obtain several amplitudes describing
interactions of pions with photons at low energies, the most famous being the
amplitude of a $\pi^0 \to \gamma \gamma$ decay~\cite{adler} (see
also~\cite{jackiw}).

An expression for the $\pi \gamma \to \pi \pi$ amplitude is given by the chiral
anomaly in the limit of small momenta of photon and pions~\cite{terentev}. The
corresponding interaction Lagrangian is
\begin{equation}
 \mathcal{L}_\pi
 = -\frac{ie}{8 \pi^2 F_\pi^3}
    \varepsilon^{\mu \nu \rho \sigma}
    F_{\mu \nu} \pi^0 \partial_\rho \pi^+ \partial_\sigma \overline{\pi^+},
 \label{pi-anomaly-coordinates}
\end{equation}
where $F_\pi = 92.2$~MeV is the $\pi \to \ell \nu$ decay constant~\cite[p.
1026]{pdg}.\footnote{In~\cite{pdg}, $f_{\pi^-} = F_\pi \sqrt{2}$ is used.} The
same expression was obtained almost simultaneously in other papers~\cite{wess,
adler-lee}.

In momentum representation,
\begin{equation}
 A_\pi =
 -h \varepsilon^{\mu \nu \rho \sigma}
    q_\mu \epsilon_\nu p_\rho k_{1 \sigma},
 \label{pi-anomaly}
\end{equation}
where $h = e / (4 \pi^2 F_\pi^3)$, and $q$, $p$, and $k_1$ are four-momenta of
the photon, initial $\pi^+$, and final $\pi^+$ respectively.

The way to check~\eqref{pi-anomaly-coordinates} and \eqref{pi-anomaly} by
studying $\pi^0$ production in a beam of charged pions which scatter coherently
in the Coulomb field of heavy nucleus was suggested in~\cite{terentev-3pi}. To
estimate the corrections to~\eqref{pi-anomaly}, it was assumed
in~\cite{terentev-3pi} that variation of the function $h$ near the threshold comes
mainly from the vector meson exchange diagrams, and the following expression was
obtained:\footnote{For $\sqrt{s} \gtrsim 4 m_\pi$, the $\rho$-meson width should
be taken into account in its $s$~channel contribution; see Fig.~4 and Eq.~(14)
in~\cite{kaiser}.}
\begin{equation}
 M_\pi = A_\pi \left\{
  1
  + \frac{2 f_{\rho \pi \gamma} f_{\rho \pi \pi}}{m_\rho^2 h}
    \left[
     \frac{s}{m_\rho^2 - s} + \frac{t}{m_\rho^2 - t} + \frac{u}{m_\rho^2 - u}
    \right]
  + \frac{e f_{\omega \gamma} f_{\omega 3 \pi}}{m_\omega^2 h}
    \frac{q^2}{m_\omega^2 - q^2}
 \right\},
 \label{pi-amplitude}
\end{equation}
where $s = (p + q)^2$, $t = (p - k_1)^2$, and $u = (q - k_1)^2$ are the
Mandelstam variables for the reaction $\pi^+ \gamma^* \to \pi^+ \pi^0$; $s + t +
u = 3 m_\pi^2 + q^2$. Subtraction is made in~\eqref{pi-amplitude} since, in the
limit $s, t, u, q^2 \to 0$, only the anomaly contribution should survive.
The equivalent photon approximation was used in~\cite{terentev-3pi} in order to
obtain the cross section of the reaction $\pi^+ \to \pi^+ \pi^0$ in the Coulomb
field of the nucleus from the cross section of the reaction $\pi^+ \gamma \to
\pi^+ \pi^0$.

Experimental verification of formulas~\eqref{pi-anomaly} and~\eqref{pi-amplitude}
is described in~\cite{antipov}. In the experiment, a 40~GeV pion beam
from the IHEP proton accelerator was used to produce neutral pions in the
Coulomb fields of C, Al and Fe nuclei. According to~\cite{antipov},
\begin{equation}
 F_{3 \pi}(0) \equiv h = 12.9 \pm 0.9 \pm 0.5 \pm 1.0 \text{ GeV}^{-3},
 \label{f3pi-experiment}
\end{equation}
while the theoretical number is
\begin{equation}
 h = \frac{e}{4 \pi^2 F_\pi^3} = 9.8 \text{ GeV}^{-3}.
\end{equation}
In expression~\eqref{f3pi-experiment}, the first error is statistical, the second
error is systematic, and the third error comes from the unknown phase of the
$\rho$ exchange contribution in~\eqref{pi-amplitude} (the $\omega$ contribution
is negligible).

Thus, in the case of pions, the anomaly saturates the considered amplitude in the
kinematics of the experiment~\cite{antipov}.\footnote{
 In~\cite{antipov}, events with $s < 10 m_\pi^2, \abs{t} < 3.5 m_\pi^2$ were
 selected. Photon virtuality varies in the following interval: $2 \cdot 10^{-3}
 \text{ GeV}^2 > -q^2 > ((s - m_\pi^2) / 2 E_\pi)^2$, $E_\pi = 40$~GeV.
}

With the inclusion of the chiral one- and two-loop corrections~\cite{bijnens,
hannah} and electromagnetic corrections~\cite{ametller}, a reanalysis of the
Serpukhov data leads to the value $F_{3 \pi} = 10.7 \pm 1.2\text{ GeV}^{-3}$;
see also~\cite{hoferichter}. The COMPASS Collaboration had plans to measure
$F_{3 \pi}$ with better accuracy~\cite[Sec. 4.2]{compass}.

\begin{figure}
 \centering
 \includegraphics{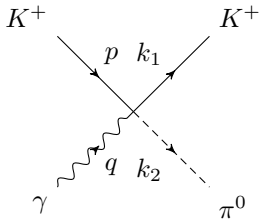}
 \caption{Momenta of particles in the $K^+ \gamma \to K^+ \pi^0$ reaction.}
 \label{anomaly-vertex}
\end{figure}

As long as the strange quark mass can be considered small in comparison with
$\Lambda_\text{QCD}$, chiral symmetry is generalized to $SU(3)_L \times SU(3)_R$
and amplitudes containing $K$~mesons can be predicted from a consideration of
the anomaly as well. This was done in~\cite{wess}, where, in particular, the
anomaly contribution to the $K^+ \gamma \to K^+ \pi^0$ amplitude was found. It
appears to be similar to~\eqref{pi-anomaly-coordinates} and
\eqref{pi-anomaly}:\footnote{
 The coefficients in~\eqref{K-anomaly-coordinates} and \eqref{K-anomaly} are
 three times bigger than in~\cite{wess} due to an extra color factor, $N_c = 3$.
}
\begin{gather}
 \mathcal{L}_K
 = - \frac{ie}{8 \pi^2 F_\pi^3}
   \varepsilon^{\mu \nu \rho \sigma}
   F_{\mu \nu} \pi^0 \partial_\rho K^+ \partial_\sigma \overline{K^+},
 \label{K-anomaly-coordinates}
 \\
 A = -\frac{e}{4 \pi^2 F_\pi^3}
     \varepsilon^{\mu \nu \rho \sigma}
     q_\mu \epsilon_\nu p_\rho k_{1 \sigma},
 \label{K-anomaly}
\end{gather}
where the particles' momenta are defined as in Fig.~\ref{anomaly-vertex}.
Analogously to the case of pions, the manifestation of anomaly~\eqref{K-anomaly}
can be looked for in $\pi^0$ production in the beam of charged kaons scattered
coherently in the Coulomb field of heavy nucleus. Such an experiment is under
way at Serpukhov~\cite{ihep}, where scattering of a $K^+$ beam with energy $E_K
= 18$~GeV in the Coulomb field of a copper (Cu) nucleus with coherent $K \pi$
production is studied.

There are two aspects in which the case of kaons differs from the case of pions.
On the one hand, since kaons are relatively heavy, we cannot approach the point
$s = t = u = q^2 = 0$ in which the anomaly dominates as closely as in the case
of the pions. On the other hand, in the case of charged kaons, there are two
reaction channels, $K^+ \gamma^* \to K^+ \pi^0$ and $K^+ \gamma^* \to K^0
\pi^+$, and only the first is influenced by the chiral anomaly~\cite{wess}.
Thus, comparing experimental data on $K^+ \pi^0$ and $K^0 \pi^+$ production
close to the threshold, one can hope to observe the effect of the anomaly.

\begin{figure}
 \centering
 \subfloat[]{
  \includegraphics{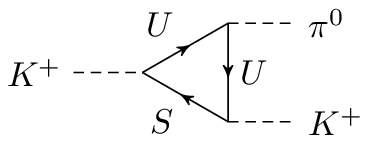}
  \label{loop-suu}
 }
 \subfloat[]{
  \includegraphics{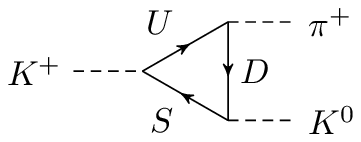}
  \label{loop-sud}
 }
 \\
 \subfloat[]{
  \includegraphics{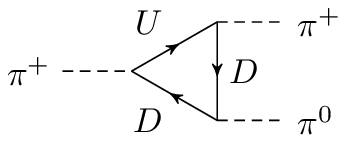}
  \label{loop-dud}
 }
 \subfloat[]{
  \includegraphics{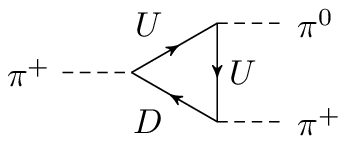}
  \label{loop-duu}
 }
 \caption{
  Attaching photon to the Pauli-Villars regulator fields, we get box diagrams
  describing anomaly contributions to (a) the $K^+ \gamma \to K^+ \pi^0$, (b)
  $K^+ \gamma \to K^0 \pi^+$ and (c), (d) $\pi^+ \gamma \to \pi^+ \pi^0$
  reactions.
 }
 \label{loops}
\end{figure}

To clarify why the amplitude of the reaction $K^+ \gamma \to K^+ \pi^0$ contains
the anomaly while that of $K^+ \gamma \to K^0 \pi^+$ is anomaly free, one should
look at Fig.~\ref{loops}, where, for completeness, diagrams for the reaction $\pi^+
\gamma \to \pi^+ \pi^0$ are shown as well.  The photon should be attached to
Pauli-Villars fields running inside the triangle diagrams drawn in
Fig.~\ref{loops} in all possible ways, leading to proportionality of the sum of
the corresponding box diagrams to the sum of electric charges of regulators. It
gives $\tfrac23 + \tfrac23 - \tfrac13 = 1$ for Fig.~\ref{loop-suu}, $\tfrac23 -
\tfrac13 - \tfrac13 = 0$ for Fig.~\ref{loop-sud}, $\tfrac23 - \tfrac13 -
\tfrac13 = 0$ for Fig.~\ref{loop-dud} and, finally, $\tfrac23 + \tfrac23 -
\tfrac13 = 1$ for Fig.~\ref{loop-duu}. Thus, the anomaly contributions to the
processes $\pi^+ \gamma \to \pi^+ \pi^0$ and $K^+ \gamma \to K^+ \pi^0$ are
equal, while that to the process $K^+ \gamma \to K^0 \pi^+$ equals zero. For an
appropriate presentation of the calculation of the anomalous amplitudes with the
help of regulator fields, see textbook~\cite[Ch. 6a]{georgi}.

We have the following plan for this paper. In Sec.~\ref{s:photonic-xsections},
formulas for the cross sections of the reactions $K^+ \gamma \to K^+ \pi^0$ and
$K^+ \gamma \to K^0 \pi^+$ at low $s$ are obtained. In
Sec.~\ref{s:nuclear-xsections}, with the help of the equivalent photon
approximation, these formulas are converted into the expressions for the cross
sections of the $K^+ N \to K^+ \pi^0 N$ and $K^+ N \to K^0 \pi^+ N$ reactions.
Comparing corresponding plots, we will conclude that extraction of the anomaly
contribution into the first reaction from the experimental data should be possible.
We summarize our results in the Conclusions. In Appendix~\ref{appendix:xsection}
we present the ``back of the envelope'' derivation of the induced by the anomaly
part of the amplitude cross section of the reaction $K^+ \gamma \to K^+ \pi^0$,
while in Appendix~\ref{appendix:couplings} numerical values of the coupling
constants used in Sec.~\ref{s:photonic-xsections} are derived.

\section{
 Cross sections of $K^+ \gamma \to K^+ \pi^0$ and $K^+ \gamma \to K^0 \pi^+$
 reactions at low invariant masses
}

\label{s:photonic-xsections}

Let us start with the calculation of the anomaly contribution to the cross
section. Momenta of particles are shown in Fig.~\ref{anomaly-vertex}, $s = (p +
q)^2$, $t = (p - k_1)^2$, and $u = (p - k_2)^2$ are Mandelstam variables, and we
neglect photon virtuality in this section, $q^2 = 0$.  The standard formula for
a differential cross section is
\begin{equation}
 \frac{d \sigma_r}{d t}
 = \frac{\overline{\abs{A}^2}}{16 \pi (s - m_{K^+}^2)^2},
 \label{anomaly-xsection}
\end{equation}
where $\overline{\abs{A}^2}$ stands for the square of the amplitude averaged
over the two transversal photon polarizations.  One should use the expression
for $A$ from~\eqref{K-anomaly}, obtaining
\begin{equation}
 \overline{\abs{A}^2}
 = \frac{e^2}{128 \pi^4 F_\pi^6}
   [ - t (s - m_{K^+}^2)^2
     - t^2 s
     + m_{\pi^0}^2 t (s + m_{K^+}^2)
     - m_{K^+}^2 m_{\pi^0}^4
   ],
 \label{anomaly-amplitude}
\end{equation}
and integrate~\eqref{anomaly-xsection} in the following interval:
\begin{multline}
   \frac{m_{\pi^0}^4}{4 s}
 - \left(
      \frac{s - m_{K^+}^2}{2 \sqrt{s}}
    + \frac{\sqrt{[s - (m_{K^+} + m_{\pi^0})^2] [s - (m_{K^+} - m_{\pi^0})^2]}}
           {2 \sqrt{s}}
   \right)^2
 < t
 \\
 < \frac{m_{\pi^0}^4}{4 s}
 - \left(
      \frac{s - m_{K^+}^2}{2 \sqrt{s}}
    - \frac{\sqrt{[s - (m_{K^+} + m_{\pi^0})^2] [s - (m_{K^+} - m_{\pi^0})^2]}}
           {2 \sqrt{s}}
   \right)^2,
\end{multline}
getting the following result:
\begin{equation}
 \sigma_r
 = \frac{\alpha}{3 \cdot 2^{10} \pi^4 F_\pi^6}
   \frac{s - m_{K^+}^2}{s^2}
   \{[s - (m_{K^+} + m_{\pi^0})^2][s - (m_{K^+} - m_{\pi^0})^2]\}^{3/2}.
 \label{sigma-r}
\end{equation}
The simplicity of the final expression clearly demonstrates that a simple
derivation of it should exist. Such a derivation is presented in
Appendix~\ref{appendix:xsection}.

\begin{figure}
 \centering
 \subfloat[$s$ channel.]{
  \includegraphics{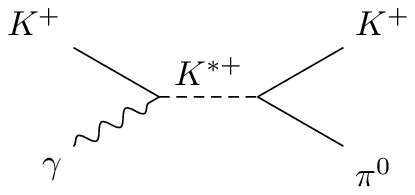}
  \label{k-pi0-s-channel}
 }
 \subfloat[$u$ channel.]{
  \includegraphics{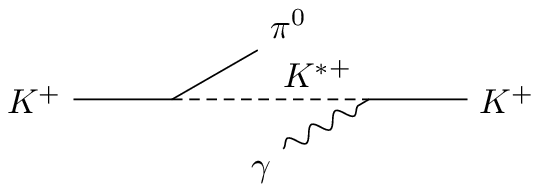}
  \label{k-pi0-u-channel}
 }
 \subfloat[$t$ channel.]{
  \includegraphics{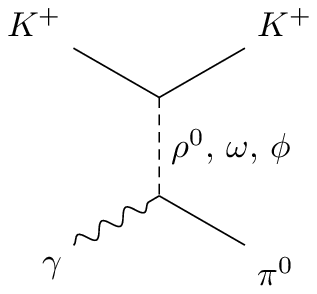}
  \label{k-pi0-t-channel}
 }
 \caption{
  Vector meson exchanges in the process $K^+ \gamma \to K^+ \pi^0$: $K^{*+}$
  meson in (a) $s$ and (b) $u$ channels, and (c) $\rho^0$, $\omega$, $\phi$
  mesons in the $t$ channel.
 }
 \label{k-pi0-diagrams}
\end{figure}

\begin{figure}
 \centering
 \subfloat[]{
  \includegraphics{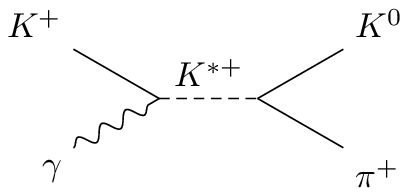}
  \label{k0-pi-s-channel}
 }
 \subfloat[]{
  \includegraphics{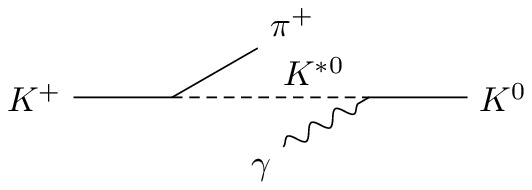}
  \label{k0-pi-u-channel}
 }
 \subfloat[]{
  \includegraphics{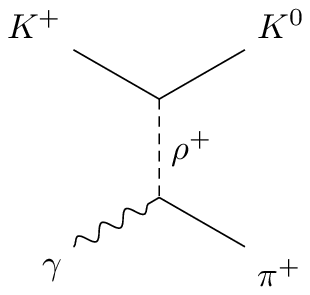}
  \label{k0-pi-t-channel}
 }
 \caption{Vector meson exchanges in the process $K^+ \gamma \to K^0 \pi^+$: (a)
 $K^{*+}$ in the $s$~channel, (b) $K^{*0}$ in the $u$~channel, (c) $\rho^+$ in
 the $t$~channel.}
 \label{k0-pi-diagrams}
\end{figure}

At low energies, the variation of the amplitude of the $K^+ \gamma \to K^+ \pi^0$
reaction comes mainly from vector meson exchange diagrams presented in
Fig.~\ref{k-pi0-diagrams}.
The exchanged mesons are $K^{*+}$ in the $s$ and
$u$~channels, and $\rho^0$, $\omega$ and $\phi$ mesons in the $t$~channel.
Similarly, for the $K^+ \gamma \to K^0 \pi^+$ reaction important meson
exchanges are $K^{*+}$ in the $s$~channel, $K^{*0}$ in the $u$~channel, and
$\rho^+$ in the $t$~channel. The corresponding diagrams are presented in
Fig.~\ref{k0-pi-diagrams}.

The amplitude for the diagram shown in Fig.~\ref{k-pi0-s-channel} is
\begin{equation}
 A_s^{(0)}(K^+ \gamma \to K^+ \pi^0)
 = \frac{2 f_{K^{*+} K^+ \gamma} f_{K^{*+} K^+ \pi^0}}
        {s - m_{K^{*+}}^2 + i \sqrt{s} \, \Gamma_{K^{*+}}(s)}
   \varepsilon^{\mu \nu \rho \sigma}
   q_\mu \epsilon_\nu p_\rho k_{1 \sigma},
\end{equation}
where $f_{K^{*+} K^+ \gamma}$ and $f_{K^{*+} K^+ \pi^0}$ are coupling constants,
\begin{equation}
 \Gamma_{K^{*+}}(s)
 = \Gamma_{K^{*+}} \frac{\sqrt{s}}{m_{K^{*+}}}
   \left[
    \frac{
     \left( 1 - \frac{(m_K - m_\pi)^2}{s} \right)
     \left( 1 - \frac{(m_K + m_\pi)^2}{s} \right)
    }{
     \left( 1 - \frac{(m_K - m_\pi)^2}{m_{K^{*+}}^2} \right)
     \left( 1 - \frac{(m_K + m_\pi)^2}{m_{K^{*+}}^2} \right)
    }
   \right]^{\frac32},
\end{equation}
and $\Gamma_{K^{*+}}$ is the total width of $K^{*+}$ as provided in~\cite{pdg}.
Since at $s, t, u \to 0$ only the anomaly contribution should survive, we
subtract from $A_s^{(0)}(K^+ \gamma \to K^+ \pi^0)$ its value at $s = 0$:
\begin{equation}
 \begin{split}
   A_s(K^+ \gamma \to K^+ \pi^0)
  &= A_s^{(0)}(K^+ \gamma \to K^+ \pi^0)
   - A_s^{(0)}(K^+ \gamma \to K^+ \pi^0) \rvert_{s = 0}
  \\
  &=     -\frac{2 f_{K^{*+} K^+ \gamma} f_{K^{*+} K^+ \pi^0}}
               {m_{K^{*+}}^2 - s - i \sqrt{s} \, \Gamma_{K^{*+}}(s)}
   \cdot \frac{s + i \sqrt{s} \, \Gamma_{K^{*+}}(s)}{m_{K^{*+}}^2}
         \varepsilon^{\mu \nu \rho \sigma}
         q_\mu \epsilon_\nu p_\rho k_{1 \sigma}.
 \end{split}
 \label{k-pi0-s-amplitude}
\end{equation}
Since $\sqrt{s} \, \Gamma_{K^{*+}}(s) \ll s$, we will neglect it in the
numerator of~\eqref{k-pi0-s-amplitude}. Performing similar calculations in the
$u$ and $t$ channels [Figs.~\ref{k-pi0-u-channel} and~\ref{k-pi0-t-channel}], we
notice that the physical regions of $t < 0$ and $u < m_{K^+}^2$ lie well below
the thresholds for decay processes of the $K^*$, $\rho$, $\omega$ and $\phi$
mesons, so no imaginary terms appear in expressions for their amplitudes.

Using the same approach for diagrams in Fig.~\ref{k0-pi-diagrams}, we obtain the
following expressions for the differential cross sections:
\begin{align}
  \frac{d \sigma(K^+ \gamma \to K^+ \pi^0)}{d t}
  &= \frac{1}{2^7 \pi}
     \left(
        t
      + \frac{(st - m_{K^+}^2 m_{\pi^0}^2) (t - m_{\pi^0}^2)}{(s - m_{K^+}^2)^2}
     \right)
  \notag \\
  &\begin{aligned}
      \times \bigg\lvert
             \frac{e}{4 \pi^2 F_\pi^3}
      &+     \frac{2 f_{K^{*+} K^+ \gamma} f_{K^{*+} K^+ \pi^0}}
                  {m_{K^{*+}}^2 - s - i \sqrt{s} \, \Gamma_{K^{*+}}(s)}
       \cdot \frac{s}{m_{K^{*+}}^2}
       +     \frac{2 f_{K^{*+} K^+ \gamma} f_{K^{*+} K^+ \pi^0}}
                  {m_{K^{*+}}^2 - u}
       \cdot \frac{u}{m_{K^{*+}}^2}
      \\
      &+     \frac{2 f_{\rho^0 \pi^0 \gamma} f_{\rho^0 K^+ K^+}}
                  {m_{\rho^0}^2 - t}
       \cdot \frac{t}{m_{\rho^0}^2}
       +     \frac{2 f_{\omega \pi^0 \gamma} f_{\omega K^+ K^+}}
                  {m_\omega^2 - t}
       \cdot \frac{t}{m_\omega^2}
       +     \frac{2 f_{\phi \pi^0 \gamma} f_{\phi K^+ K^+}}
                  {m_\phi^2 - t}
       \cdot \frac{t}{m_\phi^2}
      \bigg\rvert^2,
   \end{aligned}
  \label{k-pi0-xsection}
  \displaybreak[0] \\
  \frac{d \sigma(K^+ \gamma \to K^0 \pi^+)}{d t}
  &= -\frac{
         s t u
       - s m_{K^0}^2 m_{\pi^+}^2
       - t m_{K^+}^2 m_{K^0}^2
       - u m_{K^+}^2 m_{\pi^+}^2
       + 2 m_{K^+}^2 m_{K^0}^2 m_{\pi^+}^2
      }{2^7 \pi (s - m_{K^+}^2)^2}
  \notag \\
  &\begin{aligned}
     \times \bigg\lvert
     &\mathrel{\phantom{+}}
            \frac{2 f_{K^{*+} K^+ \gamma} f_{K^{*+} K^0 \pi^+}}
                 {m_{K^{*+}}^2 - s - i \sqrt{s} \, \Gamma_{K^{*+}}(s)}
      \cdot \frac{s}{m_{K^{*+}}^2}
      +     \frac{2 f_{K^{*0} K^0 \gamma} f_{K^{*0} K^+ \pi^+}}
                 {m_{K^{*0}}^2 - u}
      \cdot \frac{u}{m_{K^{*0}}^2}
     \\
     &-     \frac{2 f_{\rho^+ \pi^+ \gamma} f_{\rho^+ K^+ K^0}}
                 {m_{\rho^+}^2 - t}
      \cdot \frac{t}{m_{\rho^+}^2}
    \bigg\rvert^2,
   \end{aligned}
   \label{k0-pi-xsection}
\end{align}
Coupling constants $f_i$ are defined in Appendix~\ref{appendix:couplings}
with numerical values presented in Table~\ref{couplings}.
In~\eqref{k-pi0-xsection}, the most important correction comes from the $K^*$
exchange in the $s$~channel. The contribution of the $\phi$-meson exchange in
the $t$~channel is approximately 20 times smaller than the sum of the $\rho$-
and $\omega$-mesons contributions, so it could be safely neglected.
In~\eqref{k0-pi-xsection}, the term with $K^{*+}$ exchange dominates.

\begin{table}[h]
 \centering
 \caption{Coupling constants required to calculate cross sections of the $K^+
 \gamma \to K^+ \pi^0$ and $K^+ \gamma \to K^0 \pi^+$ reactions. Let us stress
 that the signs of all of the $f_{VPP}$ constants could be changed
 simultaneously, as can all the signs of $f_{VP\gamma}$, leading to two solid
 curves in Fig.~\ref{results}.}
 \begin{tabular}{|l@{$\ =\ $}D{.}{.}{4}@{\hspace{-2pt}}l|}
  \hline
  $f_{K^{*+} K^+   \pi^0}$  &  3.10  &            \\
  $f_{K^{*+} K^0   \pi^+}$  &  4.38  &            \\
  $f_{K^{*0} K^+   \pi^+}$  &  4.41  &            \\
  $f_{\rho^0 K^+   K^+}$    &  3.16  &            \\
  $f_{\rho^+ K^+   K^0}$    & -4.47  &            \\
  $f_{\omega K^+   K^+}$    &  3.16  &            \\
  $f_{\phi   K^+   K^+}$    & -4.47  &            \\
  $f_{K^{*+} K^+   \gamma}$ &  0.240 & GeV$^{-1}$ \\
  $f_{K^{*0} K^0   \gamma}$ & -0.385 & GeV$^{-1}$ \\
  $f_{\rho^0 \pi^0 \gamma}$ &  0.252 & GeV$^{-1}$ \\
  $f_{\rho^+ \pi^+ \gamma}$ &  0.219 & GeV$^{-1}$ \\
  $f_{\omega \pi^0 \gamma}$ &  0.696 & GeV$^{-1}$ \\
  $\abs{f_{\phi   \pi^0 \gamma}}$ &  0.040 & GeV$^{-1}$ \\
  \hline
 \end{tabular}
 \label{couplings}
\end{table}

\section{
 Cross sections of $K^+ N \to K^+ \pi^0 N$ and $K^+ N \to K^0 \pi^+ N$ reactions
 in the equivalent photon approximation
}

\label{s:nuclear-xsections}

\begin{figure}
 \centering
 \includegraphics{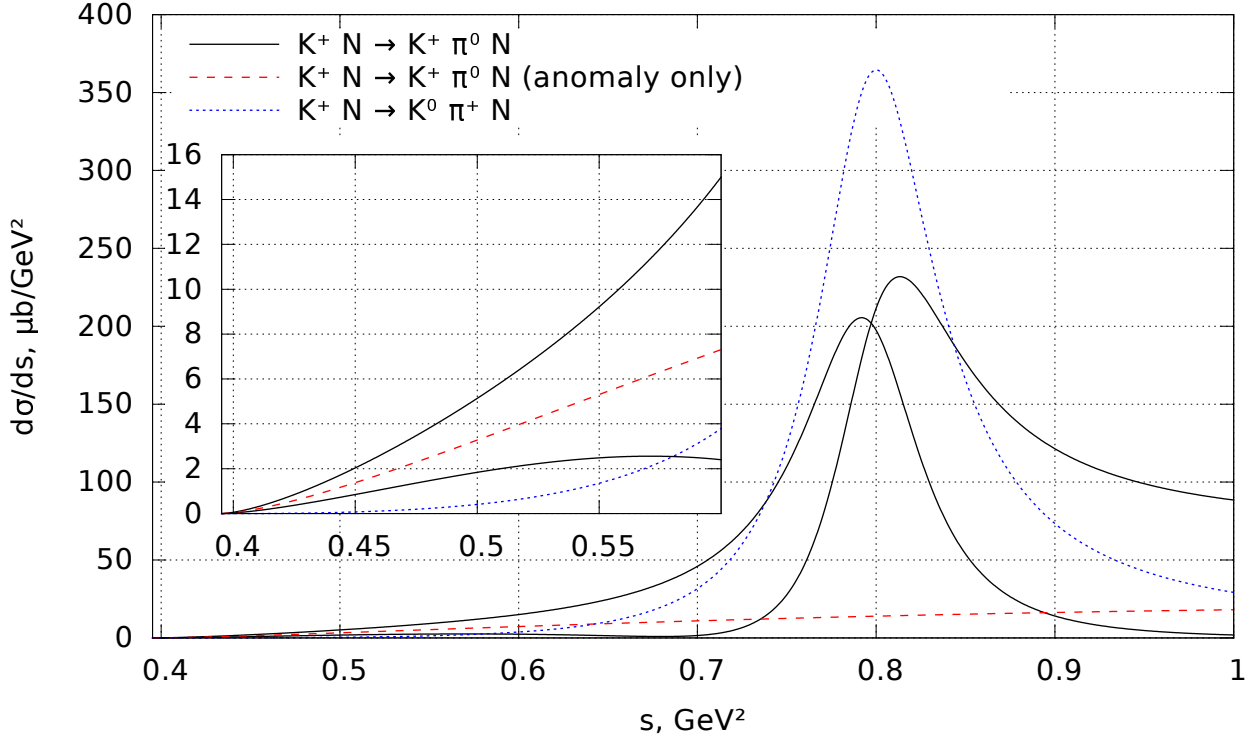}
 \caption{Differential cross sections of the reactions $K^+ N \to K^+ \pi^0 N$
 and $K^+ N \to K^0 \pi^+ N$ for $N = {}^{63}$Cu. The two solid lines for the
 $K^+ N \to K^+ \pi^0 N$ reaction correspond to the two possible choices of the
 signs of the product of coupling constants (see
 Appendix~\ref{appendix:couplings}).}
 \label{results}
\end{figure}

In the equivalent photon approximation, the following formula for the cross
section of pion production in the Coulomb field of a nucleus $N$
holds~\cite{akhiezer}:
\begin{equation}
 \frac{d \sigma(K^+ N \to K \pi N)}{dt ds dq_\perp^2}
 = \frac{Z^2 \alpha}{\pi}
   \frac{q_\perp^2}{[q_\perp^2 + (s - m_{K^+}^2)^2 / (4 E_K^2)]^2 (s - m_{K^+}^2)}
   \frac{d \sigma(K^+ \gamma \to K \pi)}{dt}
   \cdot \abs{F(\vec q\,^2)}^2,
\end{equation}
where the nucleus form factor $F(\vec q\,^2)$ is taken into account,
\begin{equation}
 F(\vec q\,^2)
 = \exp\left( -\frac{\left< r^2 \right> \vec q\,^2}{6}\right),
 \label{form-factor}
\end{equation}
$\left< r^2 \right>$ is the mean-square radius of the nucleus, $\left< r^2
\right>^{1/2} = r_0 A^{1/3}$, $r_0 = 0.94$~fm, and $A$ is the number of nucleons
in the nucleus, $A = 63$ for copper.  In~\eqref{form-factor}, $\vec q\,^2 \equiv
\vec q_\perp^{\;2} + \vec q_\parallel^{\;2}$, but, since for $s = 0.5\text{
GeV}^2$ we have 
\begin{equation*}
 a
 \equiv \frac{\left< r^2 \right> \vec q_\parallel^{\;2}}{3}
 =      \frac{\left< r^2 \right>}{3}
        \left( \frac{s - m_{K^+}^2}{2 E_K} \right)^2
 =      6 \cdot 10^{-3}
 \ll 1,
\end{equation*}
we can safely neglect $\vec q_\parallel^{\; 2}$ there.  Integration over $\vec
q_\perp^{\;2}$ from zero to infinity results in
\begin{equation}
 \frac{d \sigma(K^+ N \to K \pi N)}{dt ds}
 =     \frac{Z^2 \alpha}{\pi}
 \cdot \frac{E_1(a) - 1}{s - m_{K^+}^2}
 \cdot \frac{d \sigma(K^+ \gamma \to K \pi)}{dt},
 \label{dxs/dt-ds-dq}
\end{equation}
where $E_1(a)$ is the exponential integral,
\begin{equation}
 E_1(a) = \int_a^\infty \frac{\mathrm{e}^{-z}}{z} \, dz.
\end{equation}
In order to obtain differential cross sections $d \sigma/ds$ of the reactions
under study as functions of the invariant mass of the produced $K \pi$ system,
Eq.~\eqref{dxs/dt-ds-dq} is numerically integrated over~$t$. The result of the
integration is presented in Fig.~\ref{results}.  The effect of the anomaly can
be seen through comparison of $K^+ \pi^0$ and $K^0 \pi^+$ productions at $s
\lesssim 0.55 \text{ GeV}^2$, where the anomaly contribution is the largest.

\begin{figure}
 \centering
 \includegraphics{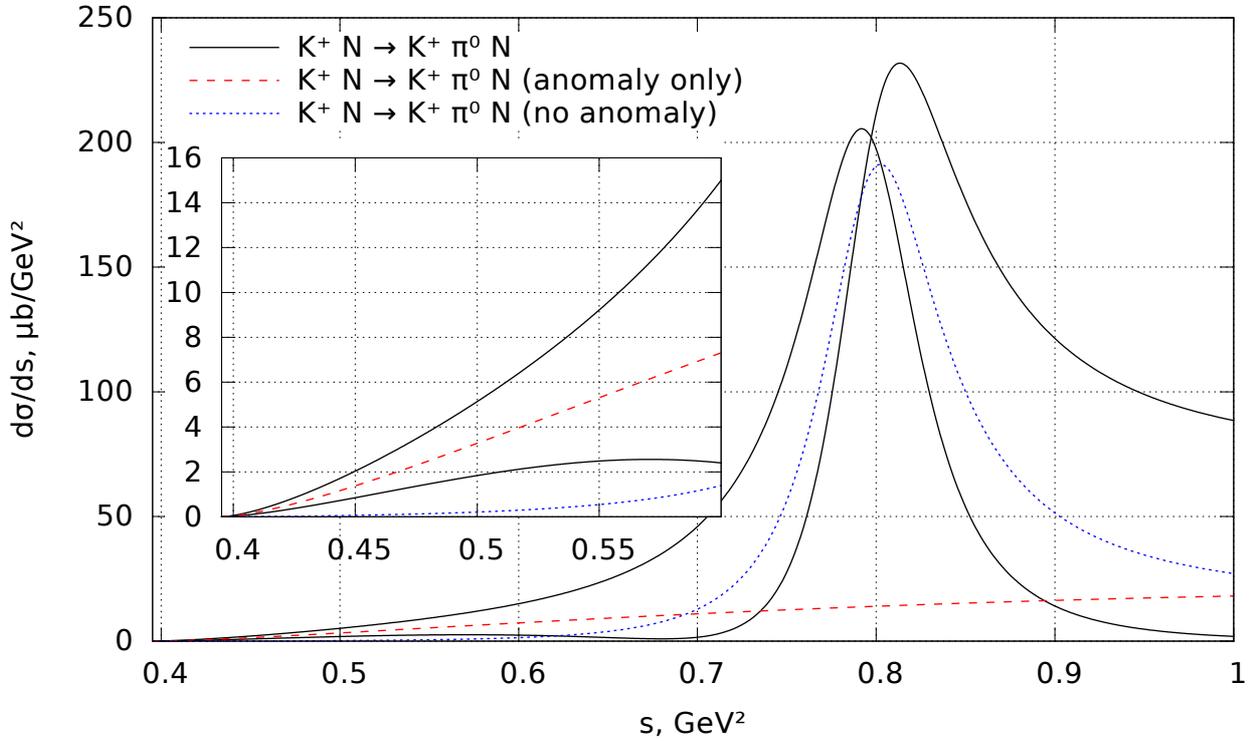}
 \caption{Differential cross sections of the reaction $K^+ N \to K^+ \pi^0 N$
 for $N = {}^{63}$Cu.  The two solid lines correspond to the two possible
 choices of the signs of the product of coupling constants.}
 \label{kplus}
\end{figure}

In Fig.~\ref{kplus}, the cross section of the reaction $K^+ N \to K^+ \pi^0 N$ is
presented and compared to that without the anomaly contribution.

\section{Conclusions}

The main results of this paper are given by
formulas~\eqref{k-pi0-xsection} and \eqref{k0-pi-xsection} and are shown in
Fig.~\ref{results}, where two solid lines for the cross section of $K^+ N \to
K^+ \pi^0 N$ reaction correspond to positive and negative signs of the product
of the coupling constants $f_i$. The luminosity of $60\ \mu\text{b}^{-1}$ is planned
to be collected in the experiment currently being performed at Serpukhov (with
the account for the detector efficiency near the $K \pi$
threshold)~\cite{obraztsov}. Integrating subplots of Fig.~\ref{results}, we get
that either 20 or 70 $K^+ \pi^0$ production events will be observed in the
interval $0.4 < s < 0.6\text{ GeV}^2$ for the destructive or constructive
interference of the anomaly and resonances terms. As for the $K^0 \pi^+$
production, about 10 events should be observed in that $s$ interval. Thus, one
can hope to observe a manifestation of the chiral anomaly in future Serpukhov
data.

Let us compare our results with those of~\cite{rogalev, burtovoy}, where the
photoproduction of pions in a charged kaon beam was studied under conditions of
the IHEP experiment as well. The main difference is that in our paper the
amplitudes describing vector meson ($K^*$, $\rho$, $\omega$, $\phi$)
contributions are subtracted at zero momenta. This subtraction is needed since
only the anomaly contribution remains at zero momenta. In~\cite{burtovoy}, only
the $K^+ \gamma \to K^+ \pi^0$ reaction is considered.  Apart from the
subtraction, our Eq.~\eqref{k-pi0-xsection} differs from Eq.~(4)
in~\cite{burtovoy} by an extra factor $-2$ in $u$ and $t$~channel
contributions.  Also, only the enhancement of the anomaly contribution by the
interference with that of the intermediate vector bosons is presented in the
figures in~\cite{burtovoy}. In~\cite{rogalev}, expressions for the cross sections
have the wrong dimension.

We are grateful to V.~F.~Obraztsov for bringing to our attention such an
interesting problem and for the numerous interesting discussions, and to
V.~A.~Novikov for the useful comments.

The authors are partially supported under RFBR Grants No.~14-02-00995 and
No.~16-02-00342. E.~Zh. is also supported by the Grant No.~MK-4234.2015.2.

\appendix
\renewcommand{\theequation}{\thesection\arabic{equation}}

\section{Simple derivation of the cross section induced by the anomaly}
\setcounter{equation}{0}
\label{appendix:xsection}

\begin{figure}
 \centering
 \includegraphics{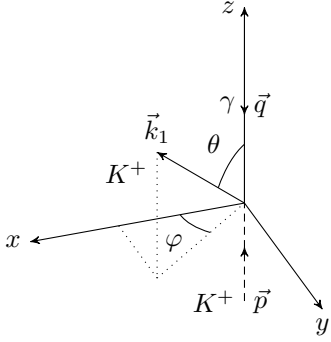}
 \caption{Kinematics of the $K^+ \gamma \to K^+ \pi^0$ reaction in the center of
 mass system. Initial particles are moving along the $z$ axis.}
 \label{cms-scattering}
\end{figure}

Let us consider the reaction $K^+ \gamma \to K^+ \pi^0$ in the center of mass
system; see Fig.~\ref{cms-scattering}. According to~\eqref{K-anomaly},
the expression for the amplitude looks like
\begin{equation}
 A = \frac{e}{4 \pi^2 F_\pi^3}
     \varepsilon^{\mu \nu \rho \sigma}
     \epsilon_\mu q_\nu p_\rho k_{1 \sigma},
\end{equation}
where $\epsilon_\mu$ is the polarization vector of the photon. Since both
$p_\rho$ and $q_\nu$ have only their third ($z$) spatial components differing
from zero, one of them should contribute by its temporal ($t$) component. As a
result, we get
\begin{equation}
 A = \frac{e}{4 \pi^2 F_\pi^3}
     \varepsilon^{\mu 0 3 \sigma}
     \epsilon_\mu k_{1 \sigma} (E_K E_\gamma + E_\gamma^2),
\end{equation}
where $E_K$ and $E_\gamma$ are the energies of the incoming $K^+$ and the photon.
When the photon polarization is parallel to the $x$ axis ($\epsilon_x =
\epsilon_1$), the $y$ component of $\vec k_1$ contributes
\begin{equation}
 A_1 = \frac{e}{4 \pi^2 F_\pi^3} 
       \abs{\vec k_1} \sin \theta \sin \varphi (E_K E_\gamma + E_\gamma^2).
\end{equation}
When the photon polarization is parallel to the $y$ axis ($\epsilon_y =
\epsilon_2$), the $x$ component of $\vec k_1$ contributes
\begin{equation}
 A_2 = \frac{e}{4 \pi^2 F_\pi^3}
       \abs{\vec k_1} \sin \theta \cos \varphi (E_K E_\gamma + E_\gamma^2).
\end{equation}
From the general formula for the differential cross section of the $2 \to 2$
reaction in the center of mass system, averaging squares of obtained amplitudes
over photon polarizations, we get
\begin{equation}
 \begin{split}
  d \sigma
  &= \frac{\overline{\abs{A(K^+ \gamma \to K^+ \pi^0)}^2}}{64 \pi^2}
     \frac{\abs{\vec p\,'}}{\abs{\vec p \,} \varepsilon^2}
     d o
  \\
  &=     \frac{\alpha}{2^8 \pi^5 F_\pi^6} \abs{\vec k_1}^2
   \cdot \frac{1}{2} \sin^2 \theta \; E_\gamma^2 (E_K + E_\gamma)^2
         \frac{\abs{\vec k_1}}{E_\gamma (E_K + E_\gamma)^2}
         d \varphi d \cos \theta.
 \end{split}
\end{equation}
Integrating the differential cross section over angles and taking into account
that in the center of mass system
$E_\gamma = \dfrac{s - m_{K^+}^2}{2 \sqrt{s}}$,
$\abs{\vec k_1}
 = \dfrac{\{[s - (m_{K^+} + m_{\pi^0})^2][s - (m_{K^+} - m_{\pi^0})^2]\}^{1/2}}
         {2 \sqrt{s}}$,
we get
\begin{equation}
 \sigma_r
 = \frac{\alpha}{3 \cdot 2^{10} \pi^4 F_\pi^6}
   \frac{s - m_{K^+}^2}{s^2}
   \{[s - (m_{K^+} + m_{\pi^0})^2] [s - (m_{K^+} - m_{\pi^0})^2]\}^{3/2},
\end{equation}
which coincides with~\eqref{sigma-r}.

\section{Coupling constants of intermediate vector bosons}
\setcounter{equation}{0}
\label{appendix:couplings}

In order to calculate cross sections through Eqs.~\eqref{k-pi0-xsection} and
\eqref{k0-pi-xsection}, we need to know the numerical values of the coupling
constants $f_i$.  These constants are defined through the following interaction
lagrangian:
\begin{equation}
 \begin{split}
  \mathcal{L_I}
  &= f_{K^{*+} K^+ \gamma}
     \varepsilon^{\mu \nu \alpha \beta}
     \partial_\mu A_\nu \partial_\alpha K^{*+}_\beta \overline{K^+}
   + i f_{K^{*+} K^+ \pi^0}
     K^{*+}_\mu
     (\overline{K^+} \partial^\mu \pi^0 - \pi^0 \partial^\mu \overline{K^+})
  \\
  &+ i f_{K^{*+} K^0 \pi^+} K^{*+}_\mu
     (\overline{K^0} \partial^\mu \overline{\pi^+}
     - \overline{\pi^+} \partial^\mu \overline{K^0})
  \\
  &+ f_{K^{*0} K^0 \gamma}
     \varepsilon^{\mu \nu \alpha \beta}
     \partial_\mu A_\nu \partial_\alpha K^{*0}_\beta \overline{K^0}
   + i f_{K^{*0} K^+ \pi^+} K^{*0}_\mu
     (\overline{K^+} \partial^\mu \pi^+ - \pi^+ \partial^\mu \overline{K^+})
  \\
  &+ f_{\rho^+ \pi^+ \gamma}
     \varepsilon^{\mu \nu \alpha \beta}
     \partial_\mu A_\nu \partial_\alpha \rho^+_\beta \overline{\pi^+}
   + i f_{\rho^+ K^+ K^0} \rho^+_\mu
     (\overline{K^+} \partial^\mu K^0 - K^0 \partial^\mu \overline{K^+})
  \\
  &+ \tfrac12 f_{\rho^0 \pi^0 \gamma}
     \varepsilon^{\mu \nu \alpha \beta}
     \partial_\mu A_\nu \partial_\alpha \rho^0_\beta \pi^0
   + \tfrac{i}{2} f_{\rho^0 K^+ K^+} \rho^0_\mu
     (K^+ \partial^\mu \overline{K^+} - \overline{K^+} \partial^\mu K^+)
  \\
  &+ \tfrac12 f_{\omega \pi^0 \gamma}
     \varepsilon^{\mu \nu \alpha \beta}
     \partial_\mu A_\nu \partial_\alpha \omega_\beta \pi^0
   + \tfrac{i}{2} f_{\omega K^+ K^+} \omega_\mu
     (K^+ \partial^\mu \overline{K^+} - \overline{K^+} \partial^\mu K^+)
  \\
  &+ \tfrac12 f_{\phi \pi^0 \gamma}
     \varepsilon^{\mu \nu \alpha \beta}
     \partial_\mu A_\nu \partial_\alpha \phi_\beta \pi^0
   + \tfrac{i}{2} f_{\phi K^+ K^+} \phi_\mu
     (K^+ \partial^\mu \overline{K^+} - \overline{K^+} \partial^\mu K^+)
  \\
  &+ \text{h.c.}
 \end{split}
 \label{lagrangian}
\end{equation}
Absolute values of some of the constants $f_i$ are obtained from the partial
widths. Other constants are deduced from these values with the help of the
$SU(3)$ symmetry between the $u$, $d$, and $s$ quarks, which also determines the
relative sings of the coupling constants.  Results of the calculations presented
below are summarized in Table~\ref{couplings}.

\begin{figure}
 \centering
 \subfloat[]{
  \includegraphics{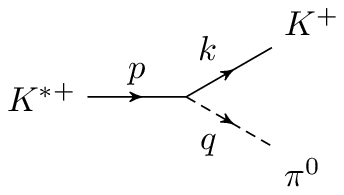}
  \label{K*-K-pi0-vertex}
 }
 \subfloat[]{
  \includegraphics{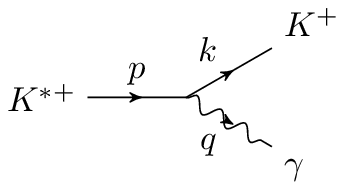}
  \label{K*-K-gamma-vertex}
 }
 \caption{Diagrams used for calculation of vector mesons coupling constants: (a)
 $f_{K^{*+} K^+ \pi^0}$, (b) $f_{K^{*+} K^+ \gamma}$.}
\end{figure}

First, consider the $K^{*+} \to K^+ \pi^0$ decay. Its amplitude equals (see
Fig.~\ref{K*-K-pi0-vertex} where momenta of particles are shown)
\begin{equation}
 A(K^{*+} \to K^+ \pi^0)
 = f_{K^{*+} K^+ \pi^0} K_\mu^{*+} (k - q)^\mu K^+ \pi^0.
\end{equation}
Its width
\begin{equation}
 \Gamma(K^{*+} \to K^+ \pi^0)
 = \frac{f_{K^{*+} K^+ \pi^0}^2 m_{K^{*+}}}{48 \pi}
   \left\{
    \left[ 1 - \left( \frac{m_{K^+} + m_{\pi^0}}{m_{K^{*+}}} \right)^2 \right]
    \left[ 1 - \left( \frac{m_{K^+} - m_{\pi^0}}{m_{K^{*+}}} \right)^2 \right]
   \right\}^{3/2}.
 \label{vss-width}
\end{equation}
Knowing the width, we can solve this equation for $\lvert f_{K^{*+} K^+ \pi^0}
\rvert$, while the sign of the coupling constant remains undetermined.

In the case of $K^{*+}$ mesons, PDG~\cite{pdg} provides the sum of the widths of
the $K^{*+} \to K^+ \pi^0$ and $K^{*+} \to K^0 \pi^+$ decays. In order to extract
$\Gamma(K^{*+} \to K^+ \pi^0)$, one has to use the relation following from
isotopic invariance
\begin{equation}
 f_{K^{*0} K^+ \pi^+} = f_{K^{*+} K^0 \pi^+} = \sqrt{2} f_{K^{*+} K^+ \pi^0},
 \label{isotopic-k}
\end{equation}
and take into account mass differences between the $K^+$ and $K^0$ mesons and
the $\pi^+$ and $\pi^0$ mesons:
\begin{align}
 \Gamma(K^{*+} \to K^+ \pi^0)
 &= 
   \frac{\Gamma(K^{*+} \to K \pi)}{
   1 + 2 \left(
    \dfrac{
       [m_{K^{*+}}^2 - (m_{K^0} + m_{\pi^+})^2]
       [m_{K^{*+}}^2 - (m_{K^0} - m_{\pi^+})^2]
    }{
       [m_{K^{*+}}^2 - (m_{K^+} + m_{\pi^0})^2]
       [m_{K^{*+}}^2 - (m_{K^+} - m_{\pi^0})^2]
    }
   \right)^\frac32
  }.
 \label{k*-k-pi-width}
\end{align}
A similar expression holds for the $K^{*0} \to K^+ \pi^-$ decay mode.

$SU(3)$ symmetry allows us to obtain the remaining $VPP$ coupling constants.
Substituting matrices of the pseudoscalar octet and the vector nonet,
\begin{equation}
 P = \begin{pmatrix}
         \tfrac{\pi^0}{\sqrt{2}} + \tfrac{\eta}{\sqrt{6}}
      &  \pi^+
      &  K^+
      \\ \overline{\pi^+}
      &  -\tfrac{\pi^0}{\sqrt{2}} + \tfrac{\eta}{\sqrt{6}}
      &  K^0
      \\ \overline{K^+}
      &  \overline{K^0}
      &  -\tfrac{2 \eta}{\sqrt{6}}
     \end{pmatrix},
 \ 
 V = \begin{pmatrix}
      \tfrac{\omega + \rho^0}{\sqrt{2}} & \rho^+            & K^{*+} \\
      \overline{\rho^+} & \tfrac{\omega - \rho^0}{\sqrt{2}} & K^{*0} \\
      \overline{K^{*+}} & \overline{K^{*0}}                 & \phi
     \end{pmatrix},
 \label{PV-matrices}
\end{equation}
into the expression for decay amplitude, we get:
\begin{equation}
 \begin{split}
  A_{VPP}
  &= f_{VPP} \tr [V_\mu (P i \partial^\mu P - i \partial^\mu P P)]
  \\
  &= i f_{VPP} \left\{
     \tfrac{1}{\sqrt{2}} K^{*+}_\mu
     (\overline{K^+} \partial^\mu \pi^0 - \pi^0 \partial^\mu \overline{K^+})
   + K^{*+}_\mu
     ( \overline{K^0} \partial^\mu \overline{\pi^+}
     - \overline{\pi^+} \partial^\mu \overline{K^0})
  \right. \\ &
   \mathrel{+} K^{*0}_\mu
     (\overline{K^+} \partial^\mu \pi^+ - \pi^+ \partial^\mu \overline{K^+})
   - \rho^+_\mu
     (\overline{K^+} \partial^\mu K^0 - K^0 \partial^\mu \overline{K^+})
  \\ &
   \mathrel{+} \tfrac{1}{2 \sqrt{2}} (\omega_\mu + \rho^0_\mu)
     (K^+ \partial^\mu \overline{K^+} - \overline{K^+} \partial^\mu K^+)
   - \tfrac12 \phi_\mu
     (K^+ \partial^\mu \overline{K^+} - \overline{K^+} \partial^\mu K^+)
   \},
  \\ &
   \mathrel{+} \text{h.c.,}
 \end{split}
\end{equation}
where only the terms entering~\eqref{lagrangian} are given. Comparing this
expression with the corresponding parts of Lagrangian~\eqref{lagrangian}, we get:
\begin{equation}
 \begin{aligned}
  f_{\rho^+ K^+ K^0} &= -f_{K^{*+} K^0 \pi^+}, &
  f_{\rho^0 K^+ K^+} &= f_{K^{*+} K^0 \pi^+} / \sqrt{2}, \\
  f_{\omega K^+ K^+} &= f_{\rho^0 K^+ K^+}, &
  f_{\phi K^+ K^+}   &= -\sqrt{2} f_{\rho^0 K^+ K^+}.
 \end{aligned}
 \label{pion-couplings}
\end{equation}

We use for $f_{\phi K^+ K^+}$ the value which follows from the direct
measurement of the corresponding width. Thus,
Eqs.~\eqref{vss-width}--\eqref{k*-k-pi-width} and \eqref{pion-couplings}
determine all of the $f_{VPP}$ values which we need, while their common sign
remains undetermined.

Next, consider the $K^{*+} \to K^+ \gamma$ decay. Its amplitude can be
represented in the following way:
\begin{equation}
 A(K^{*+} \to K^+ \gamma)
 = f_{K^{*+} K^+ \gamma}
   \varepsilon^{\mu \nu \rho \sigma}
   K^{*+}_\mu \epsilon_\nu k_\rho q_\sigma K^+,
\end{equation}
with momenta of particles shown in Fig.~\ref{K*-K-gamma-vertex}. The
numerical value of the coupling constant is determined by the width,
\begin{equation}
 \Gamma(K^{*+} \to K^+ \gamma)
 = \frac{f_{K^{*+} K^+ \gamma}^2 m_{K^{*+}}^3}{96 \pi}
   \left( 1 - \frac{m_{K^+}^2}{m_{K^{*+}}^2} \right)^3.
 \label{vsg-width}
\end{equation}

$SU(3)$ symmetry of strong interactions allows us to obtain other $V P \gamma$
coupling constants since the decay amplitudes are proportional to
\begin{equation}
 f_{V P \gamma} \sim \tr [(PV + VP) Q],
 \label{f_VPgamma}
\end{equation}
where $P$ and $V$ are defined in~\eqref{PV-matrices}, and $Q$ is the matrix of
the quark electric charges:
\begin{equation}
 Q = \begin{pmatrix}
      2/3 &    0 & 0    \\
        0 & -1/3 & 0    \\
        0 &    0 & -1/3
     \end{pmatrix}.
\end{equation}
Thus, we get
\begin{equation}
 f_{\rho^+ \pi^+ \gamma} = f_{\rho^0 \pi^0 \gamma} = f_{K^{*+} K^+ \gamma}, \ 
 f_{\omega \pi^0 \gamma} = 3 f_{K^{*+} K^+ \gamma}, \ 
 f_{K^{*0} K^0 \gamma}   = -2 f_{K^{*+} K^+ \gamma},
\end{equation}
and relative signs of the $V P \gamma$ coupling constants are fixed. Absolute
values of these constants entering Table~\ref{couplings} are determined from the
decay widths; the sign of $f_{\phi \pi^0 \gamma}$ remains undetermined. However,
the $\phi$-meson contribution to the decay amplitude can be neglected; see the
comment after Eq.~\eqref{k-pi0-xsection}.

We do not add the term proportional to $\tr [(P V - V P) Q]$
in~\eqref{f_VPgamma} since it will change the value of $f_{\rho^+ \pi^+
\gamma}$, while the value of $f_{\rho^0 \pi^0 \gamma}$ will not be changed. In
this way, the relation $f_{\rho^+ \pi^+ \gamma} = f_{\rho^0 \pi^0
\gamma}$---which follows from the fact that, in this decay, only the isoscalar
part of photon contributes---will be violated.

\end{document}